\documentclass[secnumeq,FBSmath]{FBSart}
\usepackage{epsfig}

\title{
Low Energy Proton-Deuteron Scattering with a Coulomb-Modified Faddeev Equation
}
\author{
S.\ Ishikawa\thanks{\textit{E-mail address}: ishikawa@i.hosei.ac.jp} 
}
\institute{
Department of Physics, Science Research Center, Hosei University, 
2-17-1 Fujimi, Chiyoda, Tokyo 102-8160, Japan} 
\runningtitle{
Low Energy Proton-Deuteron Scattering 
}
\runningauthor{S.\ Ishikawa}

\begin{document}
\maketitle
\begin{abstract}
A modified version of the Faddeev three-body equation to accommodate the 
Coulomb interaction, which was used in the study of three-nucleon bound 
states, is applied to the proton-deuteron scattering problem 
at energies below the three-body breakup threshold. 
A formal derivation of the equation in a time-independent scattering theory 
is given. 
Numerical results for phase shift parameters are presented to be compared 
with those of another methods and results of the phase shift analysis. 
Differential cross sections and nucleon analyzing powers are calculated 
with the effects of three-nucleon forces, and these results are compared 
with recent experimental data. 
The difference between the nucleon analyzing power in proton-deuteron 
scattering and that in neutron-deuteron scattering is discussed. 
\end{abstract}

\section{Introduction}

Recently, precise measurements of observables in proton-deuteron ($pd$) 
elastic scattering have been performed at several 
energies \cite{Hu83,Kn93,Sa94,Sh95,Sa96,Bi00,Sa00,Ca01,Br01,Er01,Wo02} 
for the details of nuclear forces, 
such as an off-shell difference in realistic nucleon-nucleon (NN) potentials 
or the evidence for three-nucleon forces (3NFs). 
For obtaining less ambiguous information from these experimental data, 
we need to calculate the corresponding observables accurately 
by solving the problem of the three-body system consisting of two protons 
and one neutron. 
As is well known, however, it is not trivial task to treat this three-body 
system because of the long-range Coulomb force acting between two protons. 
In ref.\ \cite{Sa79}, Sasakawa and Sawada proposed a modification of the 
three-body Faddeev equation \cite{Fa60} to accommodate Coulomb forces. 
The Coulomb-modified Faddeev equations were solved 
in refs.\ \cite{Sa81,Wu90,Wu93} to obtain the wave function of 
$\mathrm {}^3He$ bound state for realistic NN potentials 
with or without 3NFs, 
and were shown to work well both in its accuracy and the speed of 
the computation. 
In this paper, we apply the method to the $pd$ scattering problem and 
present some numerical results. 
The formulary aspect of the method in refs.\ \cite{Wu90,Wu93} is that 
we express the Faddeev equation as an integral equation in coordinate space, 
and then, solve the equation with an iterative method, 
which we called the Method of Continued Fractions \cite{Sa86,Is87}. 
This is in contrast with some different approaches proposed so far 
to accommodate the Coulomb force in three-nucleon (3N) continuum calculations: 
the screening and renormalization approach in momentum 
space Faddeev integral equations \cite{Al78,Be90,Al96,Al98}, 
the partial differential equation approach in configuration 
space \cite{Ch89,Fr96,Ch01}, 
the three-potential formalism with the Coulomb-Sturmian separable expansion 
method \cite{Pa97}, 
the Kohn variational method with the Pair-correlated Hyperspherical 
Harmonic (PHH) basis \cite{Ki95,Ki96}. 

Since the Coulomb-modified Faddeev equation was given heuristically 
in ref.\ \cite{Sa79},  we present a formal derivation of the equation based 
on a time-independent scattering theory in Sect.\ \ref{sec:formalism}, which 
should be useful in further development in more general few-body problem. 
The modification is performed by introducing an auxiliary potential to 
cancel the long-rangeness of the Coulomb potential. 
As will be explained in Sect.\ \ref{sec:formalism}, the cancellation works 
successfully at energies below the three-body breakup threshold.
Numerical results at those energies are thereby given in 
Sect.\ \ref{sec:results} for realistic NN forces and models of 3NF.  
Summary is given in Sect.\ \ref{sec:summary}.

\section{Three-Body Scattering with Coulomb Force}
\label{sec:formalism}

\subsection{Notations}

\begin{figure}[htb]
\centerline{\epsfig{file=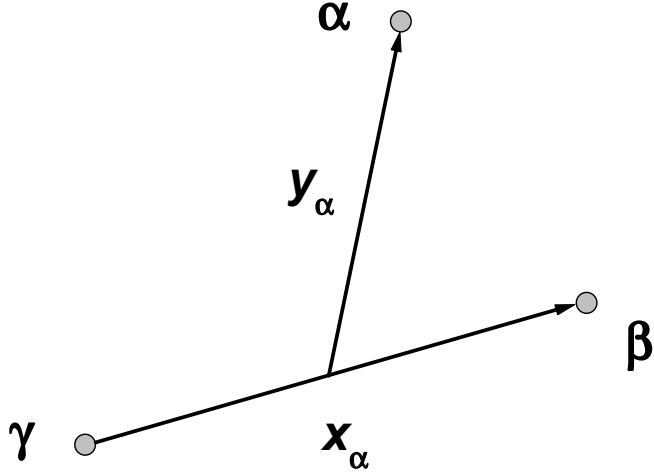,width=5in}}
\caption{The Jacobi coordinates of the three-body system.}
\label{fig:Jacobi}
\end{figure}

We consider a system of three equivalent particles (nucleons) of the mass $m$. 
The Jacobi coordinates $\{\vec{x}_\alpha,\vec{y}_\alpha\}$ are defined as 
(see Fig.\ \ref{fig:Jacobi})
\begin{equation}
\vec{x}_\alpha  =  \vec{r}_\beta - \vec{r}_\gamma, \qquad
\vec{y}_\alpha  =  \vec{r}_\alpha 
   - \frac12 \left(\vec{r}_\beta + \vec{r}_\gamma \right),
\label{eq:Jacobi}
\end{equation}
where $\alpha$, $\beta$, and $\gamma$ denote 1, 2, and 3, respectively, 
or their cyclic permutations, and $\vec{r}_\alpha$ is the position vector 
of the particle $\alpha$. 
The Hamiltonian of the system in the center-of-mass (c.m.) frame is 
\begin{equation}
H = H_0 + V = H_0 + V_1 + V_2 +  V_3, 
\end{equation}
where $H_0$ is the internal kinetic energy operator of the three-body system, 
\begin{equation}
H_0 = T_x(\vec{x}_\alpha) + T_y(\vec{y}_\alpha) 
 = -\frac{\hbar^2}{m} \nabla^2_{x_\alpha} 
 -\frac{3\hbar^2}{4m} \nabla^2_{y_\alpha}, 
\end{equation}
and $V_\alpha$ is an interaction potential for the pair $(\beta,\gamma)$, 
which consists of a short-range nuclear potential $v^s_\alpha(\vec{x}_\alpha)$ 
and the Coulomb potential $v^c_\alpha(x_\alpha)$, 
\begin{equation}
V_\alpha = v^s_\alpha(\vec{x}_\alpha) + v^c_\alpha(x_\alpha)
 = v^s_\alpha(\vec{x}_\alpha) + \frac{Z_\beta Z_\gamma e^2}{x_\alpha}.
\end{equation}
Here, $Z_\beta e$ and $Z_\gamma e$ are the electric charges of 
the particles $\beta$ and $\gamma$, respectively. 

We are going to solve a 3N Schr\"odinger equation with the c.m.\ energy $E$, 
\begin{equation}
H \Psi  = E \Psi, 
\label{eq:Sch}
\end{equation}
to obtain the scattering state $\Psi$ initiated by a state $\Xi_1$, 
\begin{equation}
\langle \vec{x}_1, \vec{y}_1 | \Xi_1 \rangle 
   = \phi^d(\vec{x}_1) F(\vec{y}_1;\vec{p}_0),
\end{equation}
where $\phi^d(\vec{x}_1)$ is the deuteron state function of the pair $(2,3)$, 
and $F(\vec{y}_1;\vec{p}_0)$ is a free state function of the nucleon 1 with 
the momentum $\vec{p}_0$. 
These functions satisfy the following Schr\"odinger equations, 
\begin{equation}
\left[ T_x(\vec{x}_1) + V_1(\vec{x}_1) \right] \phi^d(\vec{x}_1) 
  = E_d \phi^d(\vec{x}_1),
\end{equation}
and
\begin{equation}
 T_y(\vec{y}_1)  F(\vec{y_1};\vec{p}_0) = E_{p_0} F(\vec{y_1};\vec{p}_0) 
   = \frac{3\hbar^2}{4m} p_0^2 F(\vec{y}_1;\vec{p}_0),
\end{equation}
with
\begin{equation}
E  = E_d + E_{p_0}. 
\end{equation}

In a time-independent formal scattering theory, the scattering state is 
given as 
\begin{equation}
\Psi \equiv \lim_{\varepsilon \to 0 } \imath 
    \varepsilon G(E + \imath \varepsilon) \Xi_1 
  = \lim_{\varepsilon \to 0 } \Psi(\varepsilon),
\end{equation}
where $G(z)$ is the total three-body Green's function,
\begin{equation}
G(z) \equiv \frac1{z - H_0 - V}.
\end{equation}
Using the free Green's function, $G_0(z)$,
\begin{equation}
G_0(z) = \frac1{ z - H_0 },
\end{equation}
and the resolvent relation,
\begin{equation}
G(z) = G_0(z) + G_0(z) V G(z),
\label{eq:resol-G}
\end{equation}
we obtain 
\begin{equation}
\Psi(\varepsilon)  = \imath \varepsilon G_0(E+ \imath \varepsilon) \Xi_1 
  + G_0(E+ \imath \varepsilon) V \Psi(\varepsilon). 
\label{eq:Psi-eps}
\end{equation}

In the Faddeev theory with short-range potentials, the three-body state is 
divided into three components (Faddeev components),
\begin{equation}
\Psi(\varepsilon) 
  = \Phi_1(\varepsilon) + \Phi_2(\varepsilon) + \Phi_3(\varepsilon), 
\label{eq:decompose-Psi}
\end{equation}
where the Faddeev components are defined as 
\begin{equation}
\Phi_\alpha(\varepsilon) \equiv 
   \imath \varepsilon G_0(E+ \imath \varepsilon) \Xi_1~\delta_{\alpha,1}
   + G_0(E+ \imath \varepsilon)  V_\alpha  \Psi(\varepsilon). 
\label{eq:Fad-def0}
\end{equation}

Eqs.\ (\ref{eq:decompose-Psi}) and (\ref{eq:Fad-def0}) give 
\begin{equation}
\Phi_\alpha(\varepsilon) = 
\imath \varepsilon 
G_\alpha(E+ \imath \varepsilon) \Xi_1~\delta_{\alpha,1} 
 + G_\alpha(E+ \imath \varepsilon) V_\alpha 
 \left( \Phi_\beta(\varepsilon) + \Phi_\gamma(\varepsilon) \right), 
\label{eq:fad-1}
\end{equation}
where we have defined the channel Green's function $G_\alpha(z)$, 
\begin{equation}
G_\alpha (z) \equiv 
 \frac1{ z - H_0 - V_\alpha }
 = \frac1{1 - G_0(z) V_\alpha  } G_0(z). 
\end{equation}
By taking the $\varepsilon$-limit in Eq.\ (\ref{eq:fad-1}), we reach the 
Faddeev equation, 
\begin{equation}
\Phi_\alpha \equiv \lim_{\varepsilon \to 0} \Phi_\alpha(\varepsilon) 
= \Xi_1~\delta_{\alpha,1} 
 +  G_\alpha(E+ \imath 0)  V_\alpha \left( \Phi_\beta + \Phi_\gamma \right). 
\label{eq:Fad0}
\end{equation}
However, the right hand side of Eq.\ (\ref{eq:Fad0}) includes 
the following term, 
\begin{equation}
 G_0(E+ \imath 0) v^c_\alpha  (\Phi_\beta + \Phi_\gamma),
\end{equation}
which is known to cause a severe singularity due to 
the long-range Coulomb force, $v^c_\alpha(x_\alpha)$.

\subsection{The Coulomb-Modified Faddeev Equation}
\label{sec:CMF}

\begin{figure}[htb]
\centerline{\epsfig{file=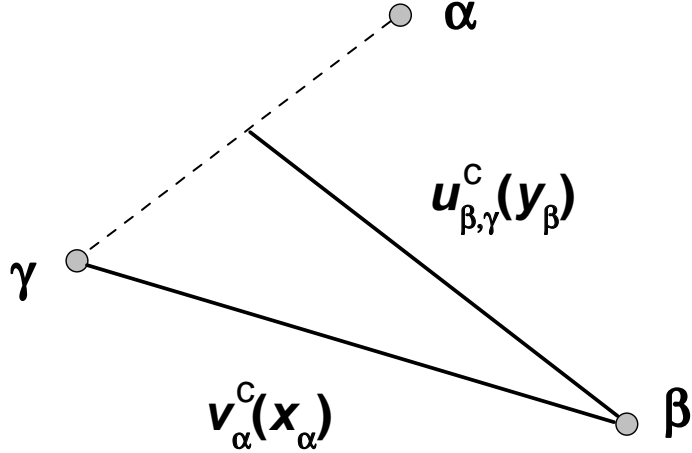,width=5in}}
\caption{
Relation between the Coulomb potential $v^c_\alpha(x_\alpha)$ and the 
auxiliary Coulomb potential $u^c_{\beta,\gamma}(y_\beta)$, 
which acts between the spectator $\beta$ and the c.m.\ of 
the pair $(\gamma,\alpha)$.} 
\label{fig:auxCoul}
\end{figure}

The singularity in the kernel of the Faddeev equation (\ref{eq:Fad0}) occurs 
due to the long-rangeness of the Coulomb interaction $v^c_\alpha(x_\alpha)$. 
We therefore consider to introduce an auxiliary potential that 
cancels the long-rangeness of $v^c_\alpha(x_\alpha)$ for a large $x_\alpha$. 
For this purpose, we define a Coulomb 
potential $u^c_{\beta,\gamma}(y_\beta)$ that 
acts between the c.m.\ of the pair $(\gamma,\alpha)$ and 
the spectator $\beta$ with respect to the charges of the pair $(\beta,\gamma)$ 
(see Fig.\ \ref{fig:auxCoul}), 
\begin{equation}
 u^c_{\beta,\gamma}(y_\beta) = \frac{Z_\beta  Z_\gamma e^2}{y_\beta}. 
\end{equation}
Using an expression, 
\begin{equation}
\vec{y}_\beta =  \vec{x}_\alpha + \frac12 \vec{x}_\beta, 
\end{equation}
which is easily given with the definition of the Jacobi 
coordinates, Eq.\ (\ref{eq:Jacobi}), we can show that 
\begin{equation}
\frac1{x_\alpha} - \frac1{y_\beta} 
  = \frac1{x_\alpha} - \frac1{|\vec{x}_\alpha + \frac12 \vec{x}_\beta|} 
    \rightarrow O(x_\alpha^{-2}), 
\end{equation} 
when $x_\alpha$  becomes large with keeping $x_\beta$ finite. 
The finiteness of $x_\beta$ is implicit in a Faddeev component $\Phi_\beta$, 
because the pair $(\gamma,\alpha)$ is in the deuteron bound state or 
in a closed channel as far as we stay 
at energies below the three-body breakup threshold. 
(See the redefinition of the Faddeev component below, 
Eq.\ (\ref{eq:Fad-def}).) 
The range of $v^c_\alpha(x_\alpha)-u^c_{\beta,\gamma}(y_\beta)$ 
is therefore short if it is combined with $\Phi_\beta$, i.e. 
\begin{equation}
\left\{v^c_\alpha(x_\alpha) - u^c_{\beta,\gamma}(y_\beta) \right\} 
 \Phi_\beta = Z_\beta Z_\gamma e^2 
   \left( \frac1{x_\alpha} - \frac1{y_\beta} \right) \Phi_\beta
  \rightarrow Z_\beta Z_\gamma e^2  O(x_\alpha^{-2}) \Phi_\beta.
\end{equation}

With this idea, we write the second term on the right hand side of 
Eq.\ (\ref{eq:Psi-eps}) as follows: 
\begin{eqnarray}
 G_0(E+ \imath \varepsilon) V \Psi(\varepsilon) &=& 
 G_0(E+ \imath \varepsilon) \biggl\{ V_1 \Phi_1(\varepsilon) + 
  V_2 \Phi_2(\varepsilon) + V_3 \Phi_3(\varepsilon) \biggr\}
\cr
 & & + G_0(E+ \imath \varepsilon) 
   \biggl\{ \left( V_1 - u^c_{2,3} \right) \Phi_2(\varepsilon) 
          + \left( V_1 - u^c_{3,2} \right) \Phi_3(\varepsilon) \biggr\} 
\cr
 & & + G_0(E+ \imath \varepsilon) 
   \biggl\{ \left( V_2 - u^c_{3,1} \right) \Phi_3(\varepsilon) 
         + \left( V_2 - u^c_{1,3} \right) \Phi_1(\varepsilon) \biggr\} 
\cr
 & & + G_0(E+ \imath \varepsilon) 
   \biggl\{ \left( V_3 - u^c_{1,2} \right) \Phi_1(\varepsilon)
         + \left(V_3 - u^c_{2,1} \right) \Phi_2(\varepsilon) \biggr\} 
\cr
 & &  + G_0(E+ \imath \varepsilon) 
   \biggl\{ u^c_1 \Phi_1(\varepsilon) + u^c_2 \Phi_2(\varepsilon) 
            + u^c_3 \Phi_3(\varepsilon) \biggr\}, 
\label{eq:G0VP}
\end{eqnarray}
where the last term on the right hand side including 
a potential $u^c_\alpha$, which is defined as a two-body Coulomb 
interaction between the pair $(\beta,\gamma)$ and the spectator $\alpha$, 
\begin{equation}
u^c_\alpha = u^c_\alpha(y_\alpha) = 
  u^c_{\alpha,\beta}(y_\alpha) + u^c_{\alpha,\gamma}(y_\alpha)
  = \frac{Z_\alpha (Z_\beta+Z_\gamma) e^2}{y_\alpha}, 
\label{eq:uhat}
\end{equation}
is added to compensate the advisedly subtracted terms 
with $u^c_{\alpha,\beta}$'s. 

We correspondingly redefine the Faddeev components 
instead of Eq.\ (\ref{eq:Fad-def0}) as 
\begin{eqnarray}
\Phi_\alpha(\varepsilon) &\equiv& 
  \imath \varepsilon G_0(E+ \imath \varepsilon)  \Xi_1~\delta_{\alpha,1} 
  + G_0(E+ \imath \varepsilon) (V_\alpha + u^c_\alpha) 
    \Phi_\alpha(\varepsilon) 
\cr
  &&   +  G_0(E+ \imath \varepsilon) 
   \left( V_\alpha - u^c_{\beta,\gamma} \right) \Phi_\beta(\varepsilon) 
\cr
  &&  +  G_0(E+ \imath \varepsilon) 
   \left( V_\alpha - u^c_{\gamma,\beta} \right) \Phi_\gamma(\varepsilon), 
\label{eq:Fad-def}
\end{eqnarray}
which gives 
\begin{eqnarray}
\Phi_\alpha(\varepsilon) &=& 
\imath \varepsilon {\cal G}_\alpha(E+ \imath \varepsilon) 
  \Xi_1~\delta_{\alpha,1}
 + {\cal G}_\alpha(E+ \imath \varepsilon) 
        \left( V_\alpha - u^c_{\beta,\gamma} \right) \Phi_\beta(\varepsilon) 
\cr
 & & +{\cal G}_\alpha(E+ \imath \varepsilon) 
    \left( V_\alpha - u^c_{\gamma,\beta} \right) \Phi_\gamma(\varepsilon), 
\label{eq:modFad-1}
\end{eqnarray}
where we have defined the channel Coulomb Green's 
function ${\cal G}_\alpha(z)$ as 
\begin{equation}
{\cal G}_\alpha (z) \equiv 
 \frac1{ z - H_0 - V_\alpha - u^c_\alpha}
 = \frac1{1 - G_0(z) (V_\alpha + u^c_\alpha) } G_0(z).
\end{equation}

Taking the $\varepsilon$-limit in Eq.\ (\ref{eq:modFad-1}) and using 
a relation, 
\begin{equation}
\lim_{\varepsilon \to 0 }  \imath \varepsilon 
 \langle \vec{x}_1, \vec{y}_1| {\cal G}_1(E+ \imath \varepsilon) 
  \Xi_1 \rangle
 = \phi^d(\vec{x}_1) F^c(\vec{y}_1;\vec{p}_0,\eta_1),
\end{equation}
where $F^c(\vec{y}_1;\vec{p}_0,\eta_1)$ is a scattering function in 
the Coulomb potential $u^c_1(y_1)$, 
\begin{equation}
\left[ T_y(\vec{y}_1) + u^c_1(y_1) \right] F^c(\vec{y}_1;\vec{p}_0,\eta_1) 
   = \frac{3\hbar^2}{4m} p_0^2 F^c(\vec{y}_1;\vec{p}_0,\eta_1),
\label{eq:Coulomb-fcn}
\end{equation}
and $\eta_\alpha$ is the Coulomb parameter, 
\begin{equation}
\eta_\alpha = \frac{2m}{3\hbar^2}\frac{Z_\alpha(Z_\beta+Z_\gamma) e^2}{p_0},
\end{equation}
we get the Coulomb-modified Faddeev equation, 
\begin{eqnarray}
\Phi_\alpha &=& \phi^d F^c(\vec{p}_0,\eta_1)~\delta_{\alpha,1}
  + {\cal G}_\alpha(E+ \imath 0) 
       \left( V_\alpha - u^c_{\beta,\gamma} \right) \Phi_\beta 
\cr
 & &   + {\cal G}_\alpha(E+ \imath 0) 
     \left( V_\alpha - u^c_{\gamma,\beta} \right) \Phi_\gamma. 
\label{eq:mod-Fad11}
\end{eqnarray}

Since the potential term $V_\alpha - u^c_{\beta,\gamma}$ is a short-range 
interaction with respect to $|\vec{x}_\alpha|$ as mentioned above, 
Eq.\ (\ref{eq:mod-Fad11}) can be treated as in the $nd$ problem. 
At the same time, it is noted that 
the appearance of the channel Coulomb Green's function 
${\cal G}_\alpha(E+ \imath 0)$ expresses that 
the spectator particle is distorted by the long-range Coulomb force. 
In the original idea of the Faddeev theory with a short-range interaction, 
while pair particles are interacting, the spectator particle is considered 
to be free. 
This picture is no more valid with the Coulomb force, and 
the auxiliary Coulomb potential creates 
the Coulomb wave function that expresses the distortion.

For the case of identical particles, the initial state should be 
\begin{equation}
\Xi = \Xi_1 + \Xi_2 + \Xi_3. 
\end{equation}
In this case, each Faddeev component has the same functional form 
and satisfies the same equation, 
\begin{equation}
\Phi_\alpha = \phi^d F^c(\vec{p}_0,\eta_\alpha) +
{\cal G}_\alpha(E+ \imath 0)
\left\{ [ V_\alpha - u^c_{\beta,\gamma} ] \Phi_\beta
      + [ V_\alpha - u^c_{\gamma,\beta} ] \Phi_\gamma \right\}.
\label{eq:mod-Fad1}
\end{equation}

\subsection{Elastic Singularity and Elastic Amplitude}

In this subsection, we derive an expression for the elastic amplitude for 
the $pd$ scattering below three-body breakup threshold. 
For this, we write the Coulomb-modified Faddeev 
equation (\ref{eq:mod-Fad1}) as 
\begin{equation}
\Phi_1 = 
 \phi^d F^c(\vec{p}_0,\eta_1)+{\cal G}_1 \Delta_1 \Phi_1, 
\label{eq:Fad}
\end{equation}
where the operator $\Delta_1$ involves a coordinate transformation, 
\begin{equation}
 \Delta_1 \Phi_1  \equiv 
\left( V_1-u^c_{2,3} \right) \Phi_2 + \left( V_1-u^c_{3,2} \right) \Phi_3. 
\label{eq:Delta}
\end{equation}

Hereafter, we drop the particle number for simplicity. 
In the case of the scattering problem, the Green's function ${\cal G}$ 
possesses a pole corresponding to the deuteron bound state. 
In order to treat this pole, we apply a subtraction method, 
in which we insert an identity, 
\begin{equation} 
1 =  |\phi^d) (\phi^d| +  \left[ 1 - |\phi^d) (\phi^d| \right], 
\end{equation} 
between ${\cal G}$ and  $\Delta$ in Eq.\ (\ref{eq:Fad}) to obtain 
\begin{equation}
\langle \vec{x}, \vec{y}| \Phi \rangle = 
  \phi^d(\vec{x}) F^c(\vec{y};\vec{p}_0,\eta) 
 + \phi^d(\vec{x}) \langle \vec{y}|\breve{G}_{c} (\phi^d| \Delta\Phi \rangle 
 + \langle \vec{x}, \vec{y}| \Gamma \Delta \Phi \rangle. 
\label{eq:el-cl}
\end{equation}
Here, $\breve{G}_{c}$ is the Coulomb Green's function for the outgoing 
particle, 
\begin{equation}
\breve{G}_{c} \equiv
  \frac1{E_{p_0} + \imath \varepsilon - T_y(\vec{y}) - u^c(y)},
\label{eq:breveG}
\end{equation}
whose analytical form is known \cite{Me65} as 
\begin{equation}
\langle \vec{y} | \breve{G}_{c} | \vec{y}^\prime \rangle 
= \sum_{\ell,m} Y_\ell^m(\hat{y}) 
  \breve{G}_{c,\ell}^{(+)}(y,y^\prime) Y_\ell^{m \ast}(\hat{y}^\prime), 
\label{eq:brG}
\end{equation}
and
\begin{equation}
\breve{G}_{c,\ell}^{(+)}(y,y^\prime) = -\frac{4m}{3\hbar^2} p_0 
 e^{\imath\sigma_\ell} \frac{u^{(+)}_\ell(p_0y_{>})}{p_0y_{>}} 
 \frac{F_\ell(p_0y_{<})}{p_0y_{<}}, 
\label{eq:brG-l}
\end{equation}
where $F_\ell(r)$ is the (real) regular Coulomb function, 
$u^{(+)}_\ell(p_0y)$ is defined with the (real) irregular Coulomb 
function $G_\ell(r)$ as 
\begin{equation}
u^{(+)}_\ell(p_0y) = 
e^{-\imath\sigma_\ell} \left( F_\ell(p_0y) + \imath G_\ell(p_0y) \right),
\end{equation}
and $\sigma_\ell$ is the Coulomb phase shift, 
\begin{equation}
\sigma_\ell = {\rm arg}\Gamma(\ell+1+\imath\eta).
\end{equation}

The operator $\Gamma$ is 
\begin{eqnarray}
\Gamma &\equiv& 
  \frac1{E + \imath \varepsilon - H_0 - V(\vec{x}) - u^c(y)} 
   \left[ 1 - |\phi^d) (\phi^d| \right]
\cr
  &=& \int d\vec{p} | F^c(\vec{p},\eta) \rangle \left\{ 
  \frac1{E_q + \imath \varepsilon - T_x(\vec{x}) - V(\vec{x})} 
 - \frac{|\phi^d)( \phi^d |}{E_q - E_d} \right\} \langle F^c(\vec{p},\eta)|, 
\cr & & 
\label{eq:Gamma}
\end{eqnarray}
where $E_q$ is the energy of the two-body subsystem, 
\begin{equation}
E_q = E - \frac{3 \hbar^2}{4m} p^2 = \frac{\hbar^2}{m} q^2, 
\label{eq:Eq}
\end{equation}
and $F^c(\vec{p},\eta)$ is the Coulomb-modified spectator state, 
which is a solution of Eq.\ (\ref{eq:Coulomb-fcn}) with the eigenvalue of 
$\frac{3 \hbar^2}{4m} p^2$. 
Because the singularities in the first and second terms on the right hand 
side in Eq.\ (\ref{eq:Gamma}) cancel each other, 
we can apply the standard quadrature to perform the $p$-integration 
in Eq.\ (\ref{eq:Gamma}) if both terms are treated together. 

The first and second terms on the right hand side of 
Eq.\ (\ref{eq:el-cl}) may be considered as a state that describes 
the elastic scattering. 
We correspondingly define a elastic function, ${\cal F}^{el}(\vec{y})$, as 
\begin{equation}
{\cal F}^{el}(\vec{y}) \equiv F^c(\vec{y};\vec{p}_0,\eta)+ 
 \langle \vec{y}|\breve{G}_{c} ( \phi^d| \Delta\Phi \rangle.
\label{eq:F-el}
\end{equation}
Substituting a formal solution of Eq.\ (\ref{eq:el-cl}) for $\Phi$ to the 
second term on the 
right hand side of Eq.\ (\ref{eq:F-el}), we obtain the following equation, 
\begin{equation}
{\cal F}^{el}(\vec{y}) = F^c(\vec{y};\vec{p}_0,\eta) + 
 \langle \vec{y}|\breve{G}_{c} ( \phi^d| \Delta
  \frac1{1-\Gamma\Delta} \phi^d{\cal F}^{el} \rangle. 
\label{eq:elF}
\end{equation}
Although the interaction term in Eq.\ (\ref{eq:elF}) is very complicated 
because of the coordinate transformation operator 
$\Delta$, this equation has the same form as 
the Lippmann-Schwinger equation for a two-body scattering problem with 
a short-range force and the Coulomb force. 
As in the case of the two-body problem, 
using the explicit form of the Green's function $\breve{G}_{c}$, 
Eqs.\ (\ref{eq:brG}) and (\ref{eq:brG-l}), 
the asymptotic form of the function $u^{(+)}_\ell(p_0y)$, 
\begin{equation}
u^{(+)}_\ell(p_0y) \vert_{y\to \infty} =
 \exp\left( +\imath (p_0y - \eta \ln 2p_0y - \ell\pi/2)\right), 
\end{equation}
and the decomposition of 
the Coulomb scattering function $F^c(\vec{y};\vec{p}_0,\eta)$, 
\begin{equation}
F^c(\vec{y};\vec{p}_0,\eta) 
= 4\pi \sum_{\ell,m} \imath^\ell Y_\ell^{m *}(\hat{p_0}) 
 Y_\ell^m(\hat{y}) e^{\imath\sigma_\ell} \frac{F_\ell(p_0y)}{p_0y}, 
\end{equation}
we obtain an asymptotic form of the elastic wave function 
${\cal F}^{el}(\vec{y})$, 
\begin{equation}
{\cal F}^{el}(\vec{y})\vert_{y\to \infty} = F^c(\vec{y};\vec{p}_0,\eta) + 
  \hat{f}^{el}(\hat{y}) \frac{e^{\imath(p_0 y - \eta \ln 2 p_0 y)}}{y}, 
\end{equation}
where $\hat{f}^{el}(\hat{y})$ is defined as 
\begin{equation}
\hat{f}^{el}(\hat{y}) = - \frac{3 \hbar^2}{4m} \frac1{4\pi} 
   \langle\overline{F^c}(p_0\hat{y}, \eta)  \phi^d| \Delta \Phi \rangle.
\end{equation}
The scattering amplitude of the elastic scattering is therefore given as 
\begin{equation}
f(\hat{y}) = f^c(\hat{y}) + \hat{f}^{el}(\hat{y}), 
\end{equation}
where $f^c(\hat{y})$ is the pure Coulomb scattering amplitude,
\begin{equation}
f^c(\hat{y}) = -\frac{\eta}{2p_0\sin^2\theta/2} 
e^{-\imath\eta\ln (\sin^2\theta/2)+2\imath\sigma_0}.
\end{equation}

\section{Numerical Results}
\label{sec:results}

\subsection{Some Remarks on Numerical Methods}

In this section, we present some results of our calculations 
to solve the Coulomb-modified Faddeev equation (\ref{eq:Fad}) for the $pd$ 
scattering problem at energies below the three-body breakup threshold. 
Although techniques used in the numerical calculations are essentially the same 
with those in the 3N bound state studies \cite{Sa81,Wu93,Sa86,Is87}, 
we give some remarks in this subsection. 

We solve the Coulomb-modified Faddeev equation (\ref{eq:Fad}) with 
an iterative method, which is called as the Method of Continued 
Fractions \cite{Sa86,Is87}. 
In most iterative methods to solve a linear integral equation, a 
$k$-th order process involves the calculation of a function $\Pi$ given by 
operating the integral kernel to a known function $\Omega$ that is produced 
from the $(k-1)$-th order solution. 
In the present case, we have
\begin{equation}
\Pi(\vec{x}, \vec{y}) \equiv 
  \langle  \vec{x}, \vec{y} | {\cal G} \Delta \Omega \rangle
 = \phi^d(\vec{x}) 
  \langle \vec{y}| \breve{G}_{c} ( \phi^d| \Delta \Omega \rangle
   + \langle  \vec{x}, \vec{y} |\Gamma \Delta \Omega  \rangle.
\label{eq:Pi-G-Xi}
\end{equation}

The first term on the right hand side of Eq.\ (\ref{eq:Pi-G-Xi}) is 
calculated numerically using the explicit expression of 
the Green's function $\breve{G}_{c}$, 
Eqs.\ (\ref{eq:brG}) and (\ref{eq:brG-l}). 

From the definition of the $\Gamma$-operator, Eq.\ (\ref{eq:Gamma}), 
the second term on the right hand side of Eq.\ (\ref{eq:Pi-G-Xi}) reads 
\begin{equation}
\langle  \vec{x}, \vec{y} |\Gamma \Delta | \Omega \rangle = 
\int d\vec{p}  F^c(\vec{y};\vec{p},\eta) 
  \left\{ \Theta(\vec{x};\vec{p}) - \phi^d(\vec{x}) C(\vec{p}) \right\},
\end{equation}
where $C(\vec{p})$ is given by 
\begin{equation}
C(\vec{p}) = \frac1{E_q - E_d} 
   \langle \phi^d  F^c(\vec{p},\eta) | \Delta \Omega \rangle, 
\end{equation}
and $\Theta(\vec{x};\vec{p})$ by 
\begin{equation}
\Theta(\vec{x};\vec{p}) =
   \frac1{E_q + \imath \varepsilon - T_x(\vec{x}) -V(\vec{x})} 
   (  F^c(\vec{p},\eta)| \Delta \Omega \rangle. 
\end{equation}
In actual numerical calculations, this $\Theta(\vec{x};\vec{p})$ is 
obtained as a solution of Schr\"odinger-like differential equation, 
\begin{equation}
\left[ E_q - T_x(\vec{x}) - V(\vec{x})\right] \Theta(\vec{x};\vec{p}) = 
   (  F^c(\vec{p},\eta)| \Delta \Omega \rangle, 
\label{eq:Sc-like}
\end{equation}
with the boundary condition that $\Theta(\vec{x};\vec{p})$ vanishes 
at a large $x$ because $E_q$ is negative in the present case.
(See Eq.\ (\ref{eq:Eq}).) 
After a decomposition of $\Theta(\vec{x};\vec{p})$ into partial waves, 
we have a set of ordinary differential equation, whose solutions can be 
obtained with the usual technique as in the two-body problem. 

For the partial-wave decomposition, we use the following 
angular-spin-isospin functions in the $J-j$ coupling scheme, 
\begin{equation}
\left|  {}^{2S+1}L_{J}, \ell_j; J_0^\pi M_0  \right)
\left| I M_I,  \frac12 \mu;  M_T  \right), 
\label{eq:J-j}
\end{equation}
where $L$ is the relative orbital angular momentum of the pair nucleons; 
$S$ the total spin of the pair; 
$J$ the total angular momentum of the pair ($\vec{J} = \vec{L} + \vec{S}$); 
$\ell$ the orbital angular momentum of the spectator nucleon; 
$j$ the sum of $\ell$ and the spin of 
the spectator ($\vec{j} = \vec{\ell} + \vec{\frac12}$); 
$J_0$ and $M_0$ the total angular momentum of the three nucleons 
($\vec{J}_0 = \vec{J}+\vec{j}$) and its third component, respectively; 
$\pi$ the parity of the system (= $(-)^{L+\ell}$); 
$I$ and $M_I$ the total isospin and its third component of the pair; 
$\mu$ the third component of the isospin of the spectator; 
$M_T$ the third component of the total isospin of the three nucleons 
($M_T = +\frac12$ for the $pd$ scattering 
and $M_T = -\frac12$ for the $nd$ scattering). 
Note that we use the isospin function in which the isotriplet pair nucleons 
are assigned as proton-proton (neutron-neutron) for $M_I = +1 (-1)$ and 
proton-neutron for $M_I = 0$. 
With this form, we take into account the mixing of 
the total isospin, $1/2$ and $3/2$, arising from the inclusion of the Coulomb 
force and/or a charge symmetry breaking (CSB) NN forces. 
Actual form of the isospin functions, whose phases are chosen 
for convenience' sake in the coordinate transformation, is shown in 
Table \ref{tab:isospin} \cite{Sa81,Wu93}. 

In the present calculation, 3N partial-wave states for which NN force acts 
are restricted to those with total two-nucleon angular momenta $J \le 3$ 
(see Subsect.\ \ref{sec:PSP}), and the total 3N angular momentum $J_0$ 
is truncated at $J_0=19/2$ in calculating scattering observables 
in Subsects.\ \ref{sec:DCS} and \ref{sec:NVAP}. 

\begin{table}[tbh]
\caption{Three-nucleon isospin functions for $M_T = \pm \frac12$.}
\label{tab:isospin}
\begin{center}
\begin{tabular}{lcccr}
\firsthline
    & $I$ & $M_I$ & $\mu$ & $|I M_I, \frac12 \mu; M_T )$ \\
\midhline
\#0 & 0 & 0      & $\pm \frac12$ & $ \pm |00) |\frac12 \pm \frac12)$ \\
\#1 & 1 & 0      & $\pm \frac12$ & $-|10) |\frac12 \pm \frac12)$ \\
\#2 & 1 & $\pm1$ & $\mp \frac12$ & $|1 \pm1) |\frac12 \mp \frac12)$ \\
\lasthline
\end{tabular}
\end{center}
\end{table}

\subsection{Phase Shift Parameters}
\label{sec:PSP}

As a first step, we use the Malfliet-Tjon (MT) I-III NN potential \cite{Ma69} 
with parameters given in ref.\ \cite{Pa80}. 
Because of the simplicity that the MT I-III potential acts only on 
the spin triplet and the spin singlet $S$-wave NN states, 
the $S$-wave nucleon-deuteron ($Nd$) elastic 
channel is classified only by the channel spin: the sum of the nucleon spin 
and the deuteron spin, 
namely, $1/2$ (the doublet state) or $3/2$ (the quartet state). 
In Table\ \ref{tab:mt13}, calculated $S$-wave phase shifts of 
the doublet (${}^2\delta$) and the quartet (${}^4\delta$) states 
for $Nd$ elastic scattering at some energies below 
the three-body breakup threshold are compared with the results of 
the configuration-space Faddeev calculations \cite{Ch89}, 
whose results for a realistic NN potential model, 
the Argonne V$_{14}$ model (AV14) \cite{Wi84},  were published as 
benchmarks \cite{Ki01} in the $pd$ calculation. 

The differences between our calculations and those of ref.\ \cite{Ch89} are 
an order of 1\% at every energies.

\begin{table}[htb]
\caption{The $nd$ and $pd$ $S$-wave phase shifts with for the MT I-III 
potential. 
The phase shifts are in degrees.}
\label{tab:mt13}
\begin{center}
\begin{tabular}{lcccc}
\firsthline
 & \multicolumn{2}{c}{${}^2\delta_{nd}$}
 & \multicolumn{2}{c}{${}^4\delta_{nd}$}
\\
 & \multicolumn{2}{c}{-----------------------------} 
 & \multicolumn{2}{c}{-----------------------------} 
\\
$E_n^\mathrm{Lab}$ (MeV) & 
   Ref.\ \protect{\cite{Ch89}} & This work & 
   Ref.\ \protect{\cite{Ch89}} & This work \\
\midhline
0.0015  & -0.230 & -0.228 &  -2.09 & -2.09\\
0.15    & -3.28  & -3.26  & -20.4  & -20.5\\
1.5     & -20.7  & -20.6  & -55.8  & -55.9\\ 
2.45    & -28.6  & -28.6  & -66.7  & -66.7\\
3.27    & -33.6  & -33.4  & -73.6  & -73.5\\
\midhline
 & \multicolumn{2}{c}{${}^2\delta_{pd}$}
 & \multicolumn{2}{c}{${}^4\delta_{pd}$}
\\
 & \multicolumn{2}{c}{-----------------------------} 
 & \multicolumn{2}{c}{-----------------------------} 
\\
$E_p^\mathrm{Lab}$ (MeV) &
   Ref.\ \protect{\cite{Ch89}} & This work & 
   Ref.\ \protect{\cite{Ch89}} & This work \\
\midhline
0.15   & -0.537 & -0.527 &   -7.46 &  -7.44\\
1.0    &  -10.6 & -10.5  &  -37.7  & -37.2\\
1.5    & -16.2  & -16.2  &  -46.5  & -46.4\\
2.0    &  -21.1 & -21.0  &  -53.5  & -53.3\\
3.0    & -28.8  & -28.7  &  -63.8  & -63.4\\
\lasthline
\end{tabular}
\end{center}
\end{table}

Next, we examine our calculations with a more realistic NN potential, 
the Argonne V$_{18}$ model (AV18) \cite{Wi95}. 
We follow the convention of the phase shifts and mixing parameters 
in ref.\ \cite{Hu95}, where the $Nd$ elastic channel is denoted 
as ${}^{2K+1}\ell_{J_0}$ with $K$ being the channel spin 
($\vec{K} = \vec{1}+\vec{\frac{1}{2}}$, $\vec{J}_0=\vec{K}+\vec{\ell}$). 
These parameters are calculated with partial-wave amplitudes in the 
channel spin scheme, which are transformed from our partial-wave amplitudes 
in the $J-j$ coupling scheme in a trivial manner \cite{Hu95}. 
In Tables \ref{tab:av18_1MeV} and \ref{tab:av18_3MeV}, 
we present results of the phase shift parameters up to $J_0 = 5/2$ 
at $E_N^\mathrm{Lab}=1$ MeV and $E_N^\mathrm{Lab}=3$ MeV, respectively. 
In the tables, our calculations with 
$J \le 2 (3) $ are denoted as $J_\mathrm{max}=2 (3)$. 
The results are compared with those of the variational method with 
the PHH basis \cite{Ki95,Ki96}, whose results for the AV14 potential were 
also published as the benchmarks \cite{Hu95,Ki01}. 
From these tables, we see that our calculations almost converge 
at $J_\mathrm{max}=3$ and the differences from the PHH calculations are 
again order of 1\%. 
In addition, although results are not shown in the tables, we have examined 
the convergence of the partial-wave expansion of the term 
$v^c_\alpha(x_\alpha)-u^c_{\beta,\gamma}(y_\beta)$ in Eq.\ (\ref{eq:mod-Fad1}) 
by neglecting $J=3$ NN force components $v^s_\alpha(x_\alpha)$ 
in the $J_\mathrm{max}=3$ calculation. 
The discrepancies between the results and the $J_\mathrm{max}=2$ calculations 
are much less than 1\%, which shows that the effect of 
the $J=3$ components in the term 
$v^c_\alpha(x_\alpha)-u^c_{\beta,\gamma}(y_\beta)$
is negligibly small.

Tables \ref{tab:mt13}, \ref{tab:av18_1MeV}, and \ref{tab:av18_3MeV} 
demonstrate that our formalism 
gives quite promising results in calculating the $nd$ and $pd$ 
elastic scattering at low energies. 

\begin{table}
\caption{
The $nd$ and $pd$ phase shifts parameters at $E_N^\mathrm{Lab} = 1$ MeV 
with AV18 potential. 
The PHH results are taken from ref.\ \protect{\cite{Ki96}}. 
Values of all parameters are in degrees.}
\label{tab:av18_1MeV}
\begin{center}
\begin{tabular}{llcccccc}
\firsthline
 &  & \multicolumn{3}{c}{$nd$} & \multicolumn{3}{c}{$pd$} \\
 &  & \multicolumn{3}{c}{------------------------------------------} 
    & \multicolumn{3}{c}{------------------------------------------} 
\\
 & & \multicolumn{2}{c}{This work} & PHH 
   & \multicolumn{2}{c}{This work} & PHH 
\\
 $J_0^\pi$ & & $J_\mathrm{max}=2$ & $J_\mathrm{max}=3$ & 
             & $J_\mathrm{max}=2$ & $J_\mathrm{max}=3$ & 
\\
\midhline
$\frac12^+$ & ${}^4D_{1/2}$ &-0.99 &-0.99 &-0.98 &-0.770&-0.770&-0.771\\
            & ${}^2S_{1/2}$ &-18.5 &-18.1 &-18.1 &-13.8 &-13.5 &-13.2 \\
            & $\eta_{1/2+}$ &0.875 &0.919 &0.928 &0.93 &0.98   &1.02 \\
\\                                                              
$\frac12^-$ & ${}^2P_{1/2}$ &-4.13 &-4.15 &-4.13 &-3.32 &-3.33 &-3.34 \\
            & ${}^4P_{1/2}$ &12.1  &12.1  &12.0  &8.98 &8.95   &9.19 \\
     & $\varepsilon_{1/2-}$ &3.46  &3.44  &3.47  &2.96 &2.95   &2.89 \\
\\                                                              
$\frac32^+$ & ${}^4S_{3/2}$ &-46.9 &-46.9 &-46.7 &-36.9 &-36.9 &-37.0 \\
            & ${}^2D_{3/2}$ &0.572 &0.571 &0.564 &0.432 &0.431 &0.442 \\
            & ${}^4D_{3/2}$ &-1.06 &-1.06 &-1.05 &-0.821&-0.823&-0.829\\
     & $\varepsilon_{3/2+}$ &0.607 &0.606 &0.621 &0.628 &0.613 &0.784 \\
           & $\zeta_{3/2+}$ &0.516 &0.515 &0.511 &0.493 &0.493 &0.500 \\
            & $\eta_{3/2+}$ &-0.110&-0.107&-0.107&-0.093&-0.091&-0.089\\
\\                                                              
$\frac32^-$ & ${}^4F_{3/2}$ &0.124 &0.124 &0.121 &0.095 &0.095 &0.097 \\
            & ${}^2P_{3/2}$ &-4.08 &-4.09 &-4.08 &-3.29 &-3.29 &-3.30 \\
            & ${}^4P_{3/2}$ &14.0  &14.0  &13.9  &10.5 &10.4   &10.7 \\
     & $\varepsilon_{3/2-}$ &-1.24 &-1.28 &-1.24 &-1.05 &-1.08 &-1.02 \\
           & $\zeta_{3/2-}$ &-0.173&-0.177&-0.177&-0.184&-0.187&-0.182\\
            & $\eta_{3/2-}$ &-1.05 &-1.04 &-1.04 &-0.968&-0.961&-0.932\\
\\                                                              
$\frac52^+$ & ${}^4G_{5/2}$ &-0.015&-0.015&-0.015&-0.011&-0.012&-0.011\\
            & ${}^2D_{5/2}$ &0.568 &0.567 &0.560 &0.429 &0.428 &0.439 \\
            & ${}^4D_{5/2}$ &-1.12 &-1.12 &-1.11 &-0.869&-0.871&-0.879\\
     & $\varepsilon_{5/2+}$ &-0.276&-0.265&-0.277&-0.272&-0.249&-0.364\\
           & $\zeta_{5/2+}$ &-0.273&-0.270&-0.272&-0.272&-0.270&-0.300\\
            & $\eta_{5/2+}$ &-0.810 &-0.818 &-0.821 &-0.818&-0.824&-0.943\\
\\                                                              
$\frac52^-$ & ${}^4P_{5/2}$ &13.3  &13.2  &13.2  &9.9 &  9.8   &10.0 \\
            & ${}^2F_{5/2}$ &-0.064&-0.064&-0.063&-0.049&-0.049&-0.049\\
            & ${}^4F_{5/2}$ &0.129 &0.129 &0.127 &0.099 &0.099 &0.100 \\
     & $\varepsilon_{5/2-}$ &0.441 &0.443 &0.447 &0.414 &0.419 &-0.218\\
           & $\zeta_{5/2-}$ &0.383 &0.390 &0.390 &0.385 &0.391 &0.384 \\
            & $\eta_{5/2-}$ &-0.120&-0.123&-0.123&-0.120&-0.122&-0.127\\
\lasthline
\end{tabular}
\end{center}
\end{table}

\begin{table}
\caption{
The $nd$ and $pd$ phase shifts parameters at $E_N^\mathrm{Lab} = 3$ MeV 
with AV18 potential. 
The PHH results are taken from refs.\ \protect{\cite{Ki95,Ki96}}.
Values of all parameters are in degrees.}
\label{tab:av18_3MeV}
\begin{center}
\begin{tabular}{llcccccc}
\firsthline
 &  & \multicolumn{3}{c}{$nd$} & \multicolumn{3}{c}{$pd$} \\
 &  & \multicolumn{3}{c}{------------------------------------------} 
    & \multicolumn{3}{c}{------------------------------------------} 
\\
 & & \multicolumn{2}{c}{This work} & PHH 
   & \multicolumn{2}{c}{This work} & PHH 
\\
 $J_0^\pi$ & & $J_\mathrm{max}=2$ & $J_\mathrm{max}=3$ & 
             & $J_\mathrm{max}=2$ & $J_\mathrm{max}=3$ & 
\\
\midhline
$\frac12^+$ & ${}^4D_{1/2}$ & -3.86 & -3.87 & -3.85 & -3.55 & -3.56& -3.56 \\
            & ${}^2S_{1/2}$ & -35.8 & -35.2 & -35.3 & -32.8 & -32.3& -32.2 \\
            & $\eta_{1/2+}$ & 1.03  & 1.10  & 1.12  & 1.02  & 1.07 & 1.10  \\
&&&&&&&\\                                                                 
$\frac12^-$ & ${}^2P_{1/2}$ & -7.42 & -7.47 & -7.49 & -7.32 &-7.37 & -7.36 \\
            & ${}^4P_{1/2}$ & 24.3  & 24.2  & 24. 2 & 21.8  & 21.7 & 22.1  \\
     & $\varepsilon_{1/2-}$ & 6.70  & 6.67  & 6.68  & 5.76  & 5.74 & 5.71  \\
&&&&&&&\\                                                                 
$\frac32^+$ & ${}^4S_{3/2}$ & -70.0 & -70.1 & -69.9 & -62.9 & -63.0&-63.1  \\
            & ${}^2D_{3/2}$ & 2.39  & 2.37  & 2.36  & 2.09  & 2.07 & 2.15  \\
            & ${}^4D_{3/2}$ & -4.15 &-4.16  & -4.14 & -3.81 & -3.82& -3.83 \\
     & $\varepsilon_{3/2+}$ & 0.724 & 0.720 & 0.747 & 0.792 & 0.753& 0.800 \\
           & $\zeta_{3/2+}$ & 1.37  & 1.36  & 1.35  & 1.29  & 1.29 & 1.30  \\
            & $\eta_{3/2+}$ & -0.377& -0.363&-0.363 & -0.327&-0.320& -0.322\\
&&&&&&&\\                                                                 
$\frac32^-$ & ${}^4F_{3/2}$ & 0.931 & 0.929 & 0.920 & 0.830 & 0.828& 0.849 \\
            & ${}^2P_{3/2}$ & -7.10 & -7.12 & -7.18 & -7.08 & -7.11& -7.14 \\
            & ${}^4P_{3/2}$ & 26.1  & 25.9  & 26.0  & 24.0  & 23.9 & 24.2  \\
     & $\varepsilon_{3/2-}$ & -2.63 &-2.69  & -2.61 & -2.24 & -2.30& -2.20 \\
           & $\zeta_{3/2-}$ & -0.245& -0.261&-0.265 & -0.305&-0.317& -0.321\\
            & $\eta_{3/2-}$ & -3.57 & -3.52 & -3.52 & -3.14 & -3.10& -3.11 \\
&&&&&&&\\                                                                 
$\frac52^+$ & ${}^4G_{5/2}$ & -0.208& -0.208&-0.206 & -0.189&-0.189& -0.189\\
            & ${}^2D_{5/2}$ & 2.36  &2.34   & 2.33  & 2.07  & 2.05 &  2.13 \\
            & ${}^4D_{5/2}$ & -4.48 & -4.49 & -4.46 & -4.09 & -4.11& -4.13 \\
     & $\varepsilon_{5/2+}$ & -0.316& -0.285&-0.315 & -0.315&-0.257& -0.350\\
           & $\zeta_{5/2+}$ & -0.716& -0.693&-0.701 & -0.707&-0.688&-0.699 \\
            & $\eta_{5/2+}$ & -1.97 & -2.03 &-2.04  & -2.04 & -2.09&-2.07  \\
&&&&&&&\\                                                                 
$\frac52^-$ & ${}^4P_{5/2}$ & 26.2  & 26.1  &  26.0 &  23.8 & 23.7 & 23.9  \\
            & ${}^2F_{5/2}$ &-0.467 & -0.470&-0.466 &-0.430 &-0.434&-0.433 \\
            & ${}^4F_{5/2}$ & 0.961 & 0.957 & 0.951 & 0.862 & 0.858& 0.876 \\
     & $\varepsilon_{5/2-}$ & 0.481 & 0.495 & 0.538 & 0.412 & 0.434& 0.343 \\
           & $\zeta_{5/2-}$ & 0.898 & 0.927 & 0.926 & 0.897 &0.928 & 0.932 \\
            & $\eta_{5/2-}$ &-0.325 &-0.340 &-0.334 & -0.338&-0.349&-0.343 \\
\lasthline
\end{tabular}
\end{center}
\end{table}

Having established the accuracy of our calculations, we proceed to compare
our results with available experimental analysis.
In ref.\ \cite{Kn93}, a phase shift analysis (PSA) was performed for 
the $pd$ scattering at $E_p^\mathrm{Lab} = 3$ MeV, corresponding to 
$E_d^\mathrm{Lab} = 6$ MeV. 
Results of our calculations for the phase shift parameters are compared 
with the PSA results in Table \ref{tab:phase-3MeV}. 
The authors in ref.\ \cite{Kn93} noticed two significant differences of 
their PSA results from those of Faddeev calculations with the PEST16 
NN potential \cite{Be90}: the ${}^2S_{1/2}$ phase shift and 
the $S-D$ mixing parameter, $\eta_{1/2+}$.
The similar differences are observed for the comparison with our AV18 results 
in Table \ref{tab:phase-3MeV}. 
The discrepancy in the ${}^2S_{1/2}$ phase shift is related to 
the binding energy of the 3N bound state. 
While the experimental value of $\mathrm {}^3He$ binding energy is 7.72 MeV, 
the calculated value for AV18 is 6.91 MeV. 
To solve this discrepancy, we introduce a 3NF model by 
Brazil group (BR) \cite{Co83} based on the two-pion exchange process with 
the cut-off parameter of 700 MeV. 
Calculated $\mathrm {}^3He$ binding energy for the AV18 plus the BR-3NF 
(AV18+BR) is 7.79 MeV. 
At the same time, although not enough but a significant improvement 
is observed for the ${}^2S_{1/2}$ phase shift and the $\eta_{1/2+}$ 
parameter as shown in Table\ \ref{tab:phase-3MeV}. 

\begin{table}
\caption{
Phase shift parameters for $pd$ scattering at $E_p^\mathrm{Lab} = 3$ MeV 
calculated for AV18, AV18+BR and AV18+BR+LS1. 
Values of PSA are taken from ref.\ \protect{\cite{Kn93}}. 
Values of all parameters are in degrees.}
\label{tab:phase-3MeV}
\begin{center}
\begin{tabular}{llcccc}
\firsthline
& & PSA & AV18 & AV18+BR & AV18+BR+LS1 \\
\midhline
\multicolumn{2}{l}{$S$-wave phases:}\\
&${}^4S_{3/2}$ & -63.95$\pm$0.28 & -63.01 & -63.02 & -63.02 \\
&${}^2S_{1/2}$ & -24.87$\pm$0.35 & -32.29 & -27.35 & -27.43 \\
\multicolumn{2}{l}{$P$-wave phases:}\\
&${}^4P_{J_0}$ average& 23.37$\pm$0.11 & 23.09 & 23.29 & 23.19 \\
&${}^4P_{5/2}-{}^4P_{3/2}$ & 0.01$\pm$0.06 & -0.17 &  0.04 & 0.40 \\
&${}^4P_{3/2}-{}^4P_{1/2}$ & 2.49$\pm$0.11 & 2.17 & 1.88 & 2.18 \\
&${}^2P_{J_0}$ average& -7.11$\pm$0.24 & -7.24 & -7.24 & -7.26  \\
&${}^2P_{3/2}-{}^2P_{1/2}$ & -0.14$\pm$0.25 & 0.26 & 0.29 & 0.34 \\
\multicolumn{2}{l}{$D$-wave phases:}\\
&${}^4D_{J_0}$ average& -3.74$\pm$0.01 & -3.81 & -3.81 & -3.81 \\
&${}^4D_{7/2}-{}^4D_{5/2}$ & 0.31$\pm$0.03 & 0.37 & 0.37 & 0.38 \\
&${}^4D_{5/2}-{}^4D_{3/2}$ & -0.28$\pm$0.03 & -0.29 & -0.29 & -0.28\\
&${}^4D_{3/2}-{}^4D_{1/2}$ & -0.16$\pm$0.03 & -0.26 & -0.27 & -0.27\\
&${}^2D_{J_0}$ average& 1.99$\pm$0.09 & 2.06 & 2.06 & 2.06\\
&${}^2D_{5/2}-{}^2D_{3/2}$ & 0.0 (fixed) & -0.02 & -0.02 & -0.02\\
\multicolumn{2}{l}{$S-D$ mixing parameters:}\\
&$\eta_{1/2+}$ & 2.00$\pm$0.10&  1.07 & 1.51 & 1.50 \\
&$\zeta_{3/2+}$ & 1.20$\pm$0.04& 1.29 & 1.29 & 1.29 \\
&$\eta_{3/2+}$  & -0.28$\pm$0.03& -0.32 & -0.31 & -0.31 \\
\multicolumn{2}{l}{$P-P$ mixing parameters:}\\
&$\varepsilon_{1/2-}$ & 5.73$\pm$0.13 & 5.74 & 5.78 & 5.69 \\
&$\varepsilon_{3/2-}$ & -2.47$\pm$0.05 & -2.30 & -2.41 & -2.52 \\
\multicolumn{2}{l}{$P-F$ mixing parameters:}\\
&$\zeta_{3/2-}$ &-1.25$\pm$0.54 & -0.32 & -0.31 & -0.30 \\
&$\eta_{3/2-}$ & -3.13$\pm$0.23 & -3.10 & -3.11 & -3.12 \\
&$\zeta_{5/2-}$ & -0.62$\pm$0.20 & 0.93 & 0.92 & 0.91 \\
&$\eta_{5/2-}$ & -0.45$\pm$0.04 & -0.35 & -0.35 & -0.35 \\
\multicolumn{2}{l}{$D-D$ mixing parameters:}\\
&$\varepsilon_{3/2+}$ &  2.15$\pm$0.24 &  0.75 &  0.76 & 0.78 \\
&$\varepsilon_{5/2+}$ & -0.75$\pm$0.12 & -0.26 & -0.26 & -0.25\\
\lasthline
\end{tabular}
\end{center}
\end{table}

In addition, we observe a significant discrepancy in the splittings of 
the spin quartet $P$-wave phase shifts, $\delta_{{}^4P_{J_0}}$ with 
$J_0$ = 1/2, 3/2, and 5/2, although their average for AV18 and AV18+BR 
agrees with the PSA values. 
This difference is closely related with so called 
{\em the $A_y(\theta)$ puzzle}: 
calculated values of the nucleon analyzing power $A_y(\theta)$ with 
realistic NN potentials for energies below $\approx$ 30 MeV 
underestimate systematically from experimental values \cite{Ko87,Wi88}. 
In ref.\ \cite{Is99}, a combination of $P$-wave phase shifts, 
\begin{equation}
- 4 \delta_{{}^4P_{1/2}} - 5 \delta_{{}^4P_{3/2}} + 9 \delta_{{}^4P_{5/2}} 
 = 
4 \left( \delta_{{}^4P_{3/2}} - \delta_{{}^4P_{1/2}} \right) + 
  9 \left( \delta_{{}^4P_{5/2}} - \delta_{{}^4P_{3/2}} \right),
\label{eq:3PJ-dif}
\end{equation}
was shown to be a good measure of the height of $A_y$-peak. 
The value of Eq.\ (\ref{eq:3PJ-dif}) is 10.05$\pm$0.70 for 
the PSA values, 7.15 for AV18, and 7.88 for AV18+BR.
The values of AV18 and AV18+BR are smaller than the experimental values 
by about 30\%, which is consistent with the underestimation 
of the $A_y$-peak height at 3 MeV. (See Subsect.\ \ref{sec:NVAP}.)

Although we do not have any consensus to solve the $A_y(\theta)$ puzzle 
at the moment, one possible reason for the discrepancy might be an unknown 
three-nucleon force that has a character of spin-orbit forces. 
In ref.\ \cite{Ki99}, Kievsky introduced purely phenomenological 
spin-orbit three-nucleon potentials.
We take one of the 3NF models that has the longest range (LS1). 
The binding energy of the $\mathrm {}^3He$ for the LS1-3NF 
in addition to the BR-3NF (AV18+BR+LS1) is 7.77 MeV.
Results of the phase shift parameters for AV18+BR+LS1 are also shown 
in Table\ \ref{tab:phase-3MeV}.
The LS1-3NF gives a remarkable effect on the phase shift splittings 
of ${}^4P_{J_0}$ states.
As a result, the value of Eq.\ (\ref{eq:3PJ-dif}) for AV18+BR+LS1 is 12.32.

\subsection{Differential Cross Sections}
\label{sec:DCS}

\begin{figure}[htb]
\centerline{\epsfig{file=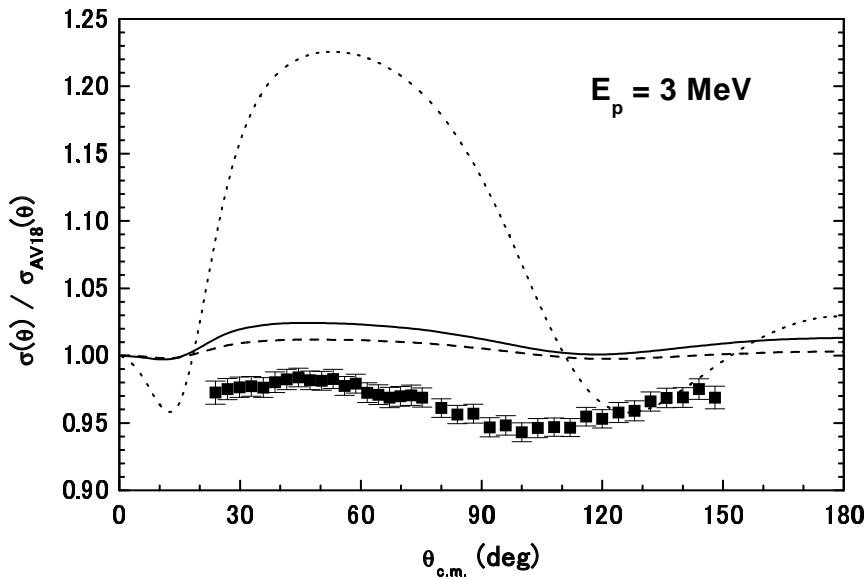,width=5in}}
\caption{
Calculated and experimental $pd$ differential cross sections normalized 
by the result of the AV18 calculation at $E_p^\mathrm{Lab} = 3.0$ MeV. 
Dotted curve denote the result of the approximated Coulomb treatment 
with the AV18 $nd$ calculation. 
Solid and dashed curves are results with the AV14$^\prime$, 
and dTRS$^\prime$ calculations, respectively.
Experimental data are taken from ref.\ \protect\cite{Sa94}.
}
\label{fig:cs-ratio30}
\end{figure}

In ref.\ \cite{Sa94}, the angular distributions of the differential 
cross section (DCS) for the $pd$ elastic scattering were measured precisely 
in the energy range of $2 \le E_p^\mathrm{Lab} \le 18$ MeV. 
In Fig.\ \ref{fig:cs-ratio30}, we plot their results 
at $E_p^\mathrm{Lab} = 3.0$ MeV being normalized to our theoretical results of 
the AV18 calculation. 
We observe that the theoretical results overestimate the experimental 
data by about 5\% at all scattering angles. 
Similar discrepancy has been observed in recent $pd$ experimental data 
at low energies \cite{Br01,Wo02}.

The authors in ref.\ \cite{Sa94} extracted the cross-section minimum, 
which appears around $\theta_\mathrm{c.m.} = 100^\circ$, 
from their data, and compared their results with the prediction of Faddeev 
calculations with the Paris NN potential, in which the $pd$ scattering 
amplitudes are approximated as the sum of the $pd$ two-body Rutherford 
amplitudes and $nd$ Faddeev amplitudes with the $pd$ two-body 
Coulomb phases as was done in ref.\ \cite{Do82}. 
As a result, they found that the ratio of the theoretical cross-section minimum 
to the experimental one reveals a strong energy dependence. 
The theoretical cross-section minimum overestimates the experimental one 
by about 20\% at $E_p^\mathrm{Lab} = 2$ MeV. 
The overestimation reduces to about 10\% at 3 MeV, 
and once disappears at 5 MeV. 
At higher energies, the calculations underestimate the experimental one. 
(See Fig. 17 in ref.\ \cite{Sa94}.) 
The dotted line in Fig.\ \ref{fig:cs-ratio30}, which denotes the result of 
AV18 with the same approximation as that in ref.\ \cite{Do82}, 
is consistent with the result of ref.\ \cite{Sa94}. 
We therefore conclude  that the overestimation of DCS at low energies 
observed in 
ref.\ \cite{Sa94} is due to the improper treatment of the Coulomb force, 
and it is reduced to about 5\% by a proper treatment of the Coulomb force. 

Besides the AV18 model, we examine some other realistic NN 
interaction models: 
a super soft-core model of de Tourreil et al. (dTRS) \cite{de75} 
and the earlier version of the Argonne model, AV14 \cite{Wi84}. 
Since the isospin-triplet component of these potentials was made as 
either proton-proton or neutron-proton force, 
a charge dependence is introduced so as to explain the difference 
among proton-proton, neutron-proton, and neutron-neutron ${}^1S_0$ scattering 
lengths \cite{Wu93}. 
The results with these models (AV14$^\prime$ and dTRS$^\prime$) are 
plotted in Fig.\  \ref{fig:cs-ratio30}.
The figure shows that the differences of the DCS 
due to the different input NN potentials observed 
at $\theta_\mathrm{c.m.} \sim 60^\circ$ 
and  $\theta_\mathrm{c.m.} \sim 180^\circ$ are at most 3\%. 
At $\theta_\mathrm{c.m.} \sim 120^\circ$, where the dependence on 
the input NN potentials is small, the calculated results overshoot 
the experimental data by about 5\%.

\begin{figure}[htb]
\centerline{\epsfig{file=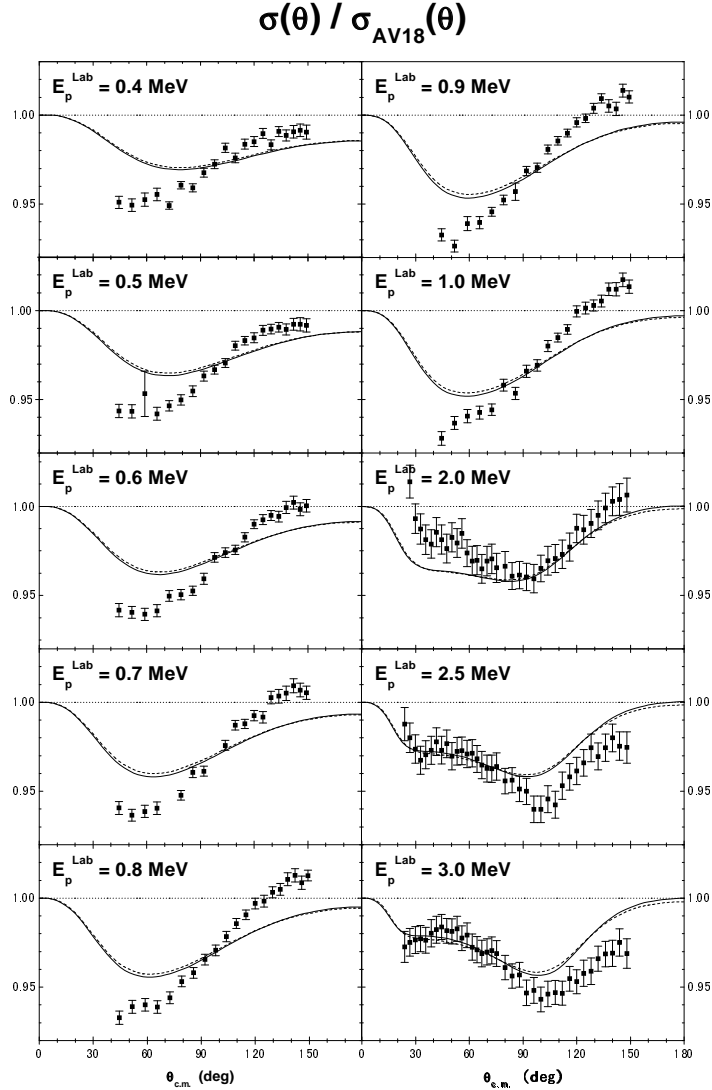,width=4.5in}}
\caption{
The $pd$ differential cross section normalized to those calculated with AV18. 
Solid curves are those with AV18+BR. 
Dashed curves are those with AV18+GS.
Experimental data are taken from ref.\ \protect\cite{Hu83} for 
$E_p^\mathrm{Lab} = 0.4 - 1.0$ MeV, and from ref.\ \protect\cite{Sa94} for 
$E_p^\mathrm{Lab} = 2.0 - 3.0$ MeV.
}
\label{fig:cs-ratio}
\end{figure}

As is seen in the results of the $pd$ $S$-wave phase shift parameters, 
we may expect that the introduction of the BR-3NF improves the DCS. 
The results of calculations with the AV18+BR normalized to 
those with the AV18 for $E_p^\mathrm{Lab} = 0.4 - 3.0$ MeV are shown as 
solid curves in Fig.\ \ref{fig:cs-ratio} together with experimental 
results of refs.\ \cite{Hu83,Sa94}. 
The figures show that the introduction of the BR-3NF improves remarkably 
the fit to the empirical cross sections. 
It is well known that the 3N binding energy and the doublet scattering 
length are strongly correlated. 
The improvement of the differential cross section due to the BR-3NF 
therefore might be nothing more than that to reproduce the 3N binding energy. 
To demonstrate this idea, we examine a fictitious spin-independent 3NF 
of the Gaussian form (GS-3NF), 
\begin{equation}
V_G=V_0^G \exp\{-(\frac{r_{21}}{r_G})^2 -(\frac{r_{31}}{r_G})^2 \} 
   + (c.p.).
\label{eq:GS-3NP}
\end{equation}
Values of the parameters, which are determined so as to reproduce 
the empirical 3N binding energy, are $r_G=1.0$ fm and $V_0^G=-40$ MeV. 
The results with the GS-3NF (AV18+GS) are shown as dashed curves 
in Fig.\ \ref{fig:cs-ratio}.
It is noted that the purely phenomenological 3NF achieves the similar 
improvement. 

\subsection{Nucleon Vector Analyzing Powers}
\label{sec:NVAP}

\begin{figure}[htb]
\centerline{\epsfig{file=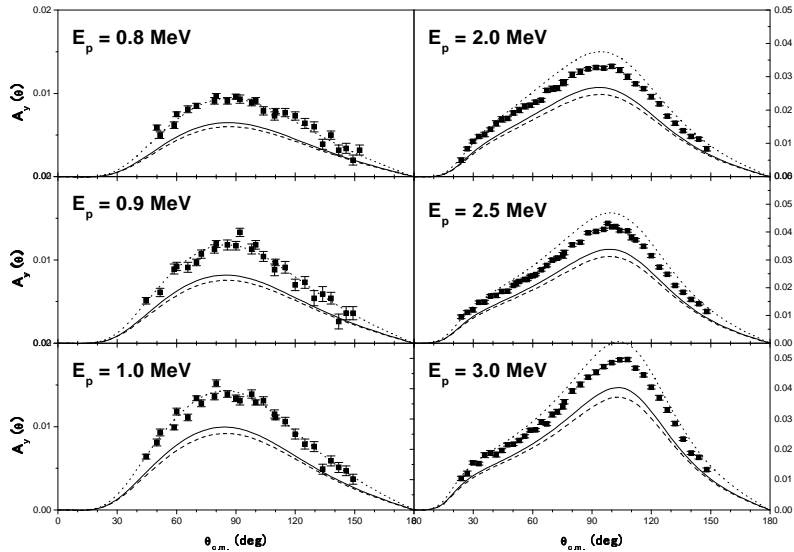,width=5in}}
\caption{
Analyzing powers of the incident proton in the $pd$ scattering 
at $E_p^\mathrm{Lab} = 0.8 - 3.0$ MeV. 
Dashed curves are calculations with AV18, solid curves those with AV18+BR, 
and dotted curves those with AV18+BR+LS1. 
Experimental date are from ref.\ \protect\cite{Hu83} for 
$E_p^\mathrm{Lab}$ = 0.8, 0.9, and 1.0 MeV, and from 
ref.\ \protect\cite{Sh95} for $E_p^\mathrm{Lab}$ = 2.0, 2.5, and 3.0 MeV.
}
\label{fig:ay}
\end{figure}

In Fig.\ \ref{fig:ay}, we show results of the proton analyzing power 
$A_y(\theta)$ at several energies below three-body breakup threshold 
comparing with available experimental data \cite{Hu83,Sh95}. 
The results with the AV18 potential (dashed curves) undershoot 
the experimental values of $A_y(\theta)$ (The $A_y$ puzzle). 
The inclusion of the BR-3NF into the nuclear Hamiltonian increases only a 
small amount of $A_y(\theta)$ as denoted by the solid curves. 
On the other hand, the phenomenological spin-orbit 3NF, LS1 \cite{Ki99}, 
improves $A_y(\theta)$ considerably as denoted by dotted curves. 
It is noted that the small amount of overestimation in $A_y(\theta)$ 
with AV18+BR+LS1 above 2 MeV is because of the  combination of 
the LS1-3NF with the BR-3NF 
since the spin-orbit 3NFs in ref.\ \cite{Ki99} are made so as to reproduce 
the experimental values of $A_y(\theta)$ by themselves, 

A merit to accommodate the Coulomb interaction in 3N system is 
that we can discuss the effect of CSB with high quality by comparing 
calculations with and without the Coulomb interaction.
In ref.\ \cite{Wu93}, CSB effects are studied for 3N bound states. 
It was demonstrated that the calculated values of the $\mathrm {}^3H$ 
binding energy ($B(\mathrm {}^3 H)$) and those of $\mathrm {}^3He$ 
($B(\mathrm {}^3He)$) for various combinations of NN forces and 3NFs 
have a linear correlation. 
When we plot the results of $B(\mathrm {}^3He)$ without any CSB effect 
(but only with the Coulomb force) against those of $B(\mathrm {}^3H)$, 
these points are fitted by a line, which does not pass through the point of 
the experimental values. 
When we include a CSB NN force, which reproduces the experimental difference 
of the proton-proton and neutron-neutron ${}^1$S$_0$ scattering lengths, 
$\Delta a$ = 1.5 fm, 
together with the higher order electromagnetic interaction, 
we can draw a line that shifts to pass the experimental point. 
The CSB force used in the study of the 3N bound states is a central one. 
On the other hand, information of non-central spin-dependent CSB forces is 
expected to be given from 
the study of polarization observables in the $nd$ and the $pd$ scatterings. 
So far, the nucleon vector analyzing power $A_y(\theta)$ was measured for 
the $nd$ and the $pd$ elastic scattering for 
some energies \cite{Sh95,Mc94,Ni98}. 
The measured values of $A_y(\theta)$ have peaks at 
$\theta_{c.m.} \sim 105^\circ$ at low energies.
The difference between the $\polarize{n}d$ and 
$\polarize{p}d$  $A_y(\theta)$ peak heights, which depends weakly on 
incident energies below $E_N^\mathrm{Lab} = 10$ MeV, and is roughly 0.01.

\begin{figure}[htb]
\centerline{\epsfig{file=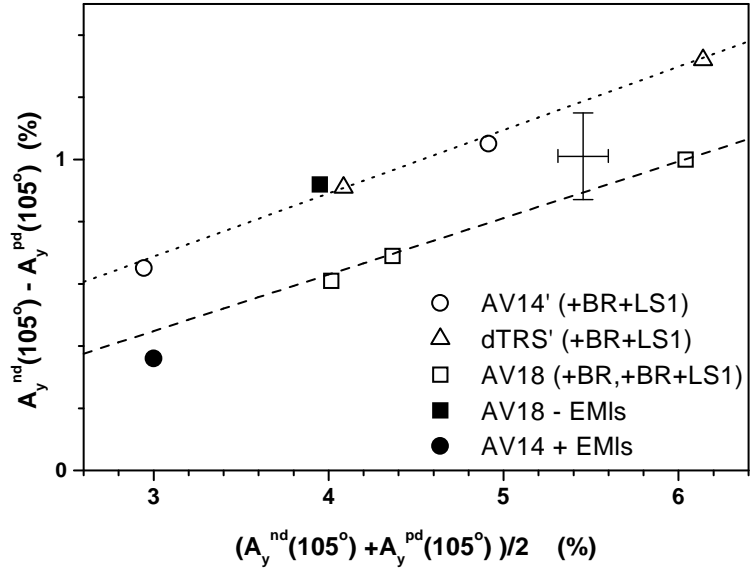,width=5in}}
\caption{
$A_y^{nd}-A_y^{pd}$ vs. $(A_y^{nd}+A_y^{pd})/2$ 
at $\theta_\mathrm{c.m.}=105^\circ$.
Experimental date are taken from \protect\cite{Sh95,Mc94}.
The meaning of the points is described in the text.
}
\label{fig:ay-nd-pd}
\end{figure}

In Fig.\ \ref{fig:ay-nd-pd}, the calculated values of the difference, 
$A_y^{nd}-A_y^{pd}|_{\theta_\mathrm{c.m.} = 105^\circ}$ 
at $E_N^\mathrm{Lab}$ = 3 MeV 
are plotted against their average values, 
$\left(A_y^{nd}+A_y^{pd}\right)/2|_{\theta_\mathrm{c.m.} = 105^\circ}$. 
In this figure, the results with AV14$^\prime$ and AV14$^\prime$+BR+LS1 
are plotted as open circles, 
those with dTRS$^\prime$ and dTRS$^\prime$+BR+LS1 as open triangles, 
and those with AV18, AV18+BR, and AV18+BR+3NF as open squares. 
Apparently, the results are divided into two groups that are 
linearly correlated: 
one is the results with AV14$^\prime$ or dTRS$^\prime$ NN potentials 
as guided by the dotted line, 
and the other is those with AV18 as guided by the dashed line. 

We remark that the difference between these lines is attributed to 
higher order electromagnetic (EM) interactions in the AV18 potential. 
The EM interactions include a spin-orbit force arising from the 
magnetic moment interaction, which contributes as a CSB force \cite{Wi95}. 
To demonstrate the effect of the EM spin-orbit CSB interaction (EMls), 
a result of calculation with AV18 neglecting EMls is plotted as 
a filled square, 
and that with AV14$^\prime$ including EMls as a filled circle 
in Fig.\ \ref{fig:ay-nd-pd}. 
The former point almost lies on the dotted line, and the latter 
on the dashed one. 
We therefore conclude that the higher order electromagnetic 
interaction in the AV18 model explains the difference of the two lines in 
Fig.\ \ref{fig:ay-nd-pd}, and gives about 10\% effect in the $nd$-$pd$ 
analyzing power difference.

In ref.\ \cite{St98}, the effect of the EM spin-orbit force is reported to 
be small in $Nd$ analyzing power. 
However, the deuteron was treated as a single spin-1 particle 
in ref. \cite{St98}. 
The present result therefore shows an importance of a three-body treatment of
the $pd$ system.

\section{Summary}
\label{sec:summary}

We have formally derived a modified Faddeev equation to accommodate the 
long-range Coulomb interaction in three-nucleon ($pd$) scattering, 
which was originally presented in ref.\ \cite{Sa79}. 
In the modification, we introduced an auxiliary Coulomb force that acts 
between a spectator proton and the c.m.\ of the rest pair nucleons 
(proton and neutron). 
This interaction takes into account two issues: 
(1) the Coulomb distortion of the spectator particle 
and (2) the cancellation of the long-rang part of the original Coulomb force 
in the integral kernel for energies below the three-body breakup threshold.
The modified Faddeev equation can be treated similarly as in the case of 
the $nd$ scattering. 

We presented some numerical results to solve the Coulomb modified 
Faddeev equation 
for the $pd$ elastic scattering at energies below the three-body 
breakup threshold for some realistic NN potential models without 
or with three-nucleon forces. 
Results of the phase shift parameters agree with the available accurate 
results by different methods. 

Differential cross sections for an NN potential overestimate the 
experimental data by about 5\% at low energies. 
This overestimation is recovered by introducing three-nucleon forces that 
reproduce the 3N binding energy. 

CSB effects are studied for the nucleon analyzing powers in 
$pd$ scattering and $nd$ scattering at 3 MeV. 
We found that a spin-orbit force due to the nucleon magnetic moment 
interaction plays a significant role in the difference.

\begin{acknowledge}
This research was supported by the Japan Society for the Promotion of Science, 
under Grant-in-Aid for Scientific Research No. 13640300.
The numerical calculations have been supported by the Computational Science 
Research Center, Hosei University, under project lab0003.
\end{acknowledge}



\begin{thebibliography}{99}

\bibitem{Hu83}
Huttel, E., et al.: Nucl. Phys. {\bf A406}, 435 (1983) 

\bibitem{Kn93}
Knutson, L. D., Lamm, L. O., McAninch, J. E.: 
Phys. Rev. Lett. {\bf 71}, 3762 (1993)

\bibitem{Sa94} 
Sagara, K., et al.: Phys. Rev. C {\bf 50}, 576 (1994)

\bibitem{Sh95}
Shimizu, S., et al.: Phys. Rev. C {\bf 52}, 1193 (1995)

\bibitem{Sa96}
Sakamoto, N., et al.: Phys. Lett. B {\bf 367}, 60 (1996)

\bibitem{Bi00}
Bieber, R., et al.: Phys. Rev. Lett. {\bf 84}, 606 (2000)

\bibitem{Sa00}
Sakai, H., et al.: Phys. Rev. Lett. {\bf 84}, 5288 (2000)

\bibitem{Ca01}
Cadman, R. V., et al.: Phys. Rev. Lett. {\bf 86}, 967 (2001)

\bibitem{Br01}
Brune, C. R., et al.: Phys. Rev. C {\bf 63}, 044013 (2001)

\bibitem{Er01}
Ermisch, K., et al.: Phys. Rev. Lett. {\bf 86}, 5862 (2001)

\bibitem{Wo02}
Wood, M. H., et al.: Phys. Rev. C {\bf 65}, 034002 (2002)

\bibitem{Sa79}
Sasakawa, T., Sawada, T.: Phys. Rev. C {\bf 20}, 1954 (1979)

\bibitem{Fa60}
Faddeev, L. D.: Zh. Eksp. Teor. Fiz. {\bf 39}, 1459 (1960)
[Sov. Phys.-JETP {\bf 12}, 1041 (1961)]

\bibitem{Sa81}
Sasakawa, T., Okuno, H., Sawada, T.: Phys. Rev. C {\bf 23}, 905 (1981)

\bibitem{Wu90}
Wu, Y., Ishikawa, S., Sasakawa, T.:
Phys. Rev. Lett. {\bf 64}, 1875 (1990); {\bf 66}, 242 (1991)

\bibitem{Wu93}
Wu, Y., Ishikawa, S., Sasakawa, T.: Few-Body Syst. {\bf 15}, 145 (1993)

\bibitem{Sa86} 
Sasakawa, T., Ishikawa, S.: Few-Body Syst. {\bf 1}, 3 (1986)

\bibitem{Is87} 
Ishikawa, S.: Nucl. Phys. {\bf A463}, 145c (1987)

\bibitem{Al78}
Alt, E. O., Sandhas, W., Ziegelmann, H.:
Phys. Rev. C {\bf 17}, 1981 (1978)

\bibitem{Be90} 
Berthold, G. H., Stadler, A., Zankel, H.:
Phys. Rev. C {\bf 41}, 1365 (1990)

\bibitem{Al96}
Alt, E. O., Sandhas, W.: 
In:  Coulomb Interactions in Nulcear and Atomic Few-Body Collisions 
(Levin, F. S.,  Micha, D. A., eds.): p. 1. 
New York and London: Plenum Press 1996

\bibitem{Al98}
Alt, E.O., Mukhamedzhanov, A. M., Sattarov, A. I.:
Phys. Rev. Lett. {\bf 81}, 4820 (1998)

\bibitem{Ch89}
Chen, C. R., et al.: 
Phys. Rev. C {\bf 39}, 1261 (1989)

\bibitem{Fr96}
Friar, J. L., Payne, G. L.: 
In:  Coulomb Interactions in Nulcear and Atomic Few-Body Collisions 
(Levin, F. S.,  Micha, D. A., eds.): p. 97. 
New York and London: Plenum Press 1996

\bibitem{Ch01}
Chen, C. R., Friar, J. L., Payne, G. L.: 
Few-Body Syst. {\bf 31}, 13 (2001)

\bibitem{Pa97} 
Papp, Z.: Phys. Rev. C {\bf 55}, 1080 (1997)


\bibitem{Ki95}
Kievsky, A., Viviani, M., Rosati, S.:
Phys. Rev. C {\bf 52}, R15 (1995)

\bibitem{Ki96} 
Kievsky, A., et al.:
Nucl Phys. {\bf A 607}, 402 (1996)


\bibitem{Me65}
Messiah, A.: Quantum Mechanics. New York: John Wiley 1965

\bibitem{Ma69}
Malfliet, R. A., Tjon, J. A.: Nucl. Phys. {\bf A127}, 161 (1969)
\bibitem{Pa80}
Payne, G. L., et al.:
Phys. Rev. C {\bf 32}, 823 (1980)

\bibitem{Wi84}
Wiringa, R. B., Smith, R. A., Ainsworth, T. L.:
Phys. Rev. C. {\bf 29}, 1207 (1984)

\bibitem{Ki01}
Kievsky A., et al.:
Phys. Rev. C {\bf 63}, 064004 (2001)



\bibitem{Wi95} 
Wiringa, R. B., Stoks, V. G. J., Schiavilla, R.:
Phys. Rev. C {\bf 51}, 38 (1995)

\bibitem{Hu95}
H\"uber, D., et al..:
Phys. Rev. C {\bf 51}, 1100 (1995)

\bibitem{Co83} 
Coelho, H. T., Das, T. K., Robilotta, M. R.:
Phys. Rev. C {\bf 28}, 1812 (1983)

\bibitem{Ko87}
Koike, Y., Haidenbauer, J.: 
Nucl. Phys. {\bf A463}, 365c (1987)

\bibitem{Wi88}
Wita\l a, H., Gl\"ockle, W., Cornelius, T.: 
Nucl. Phys. {\bf A491}, 157 (1988)

\bibitem{Is99}
Ishikawa, S.: Phys. Rev. C {\bf 59}, R1247 (1999)

\bibitem{Ki99}
Kievsky, A.: Phys. Rev. C {\bf 60}, 034001 (1999)

\bibitem{Do82}
Doleschall, P., et al.: Nucl. Phys. {\bf A380}, 72 (1982).

\bibitem{de75}
de Tourreil, R., Rouben, B., Sprung, D. W. L.:
Nucl. Phys. {\bf 242}, 445 (1975)

\bibitem{Mc94} 
McAninch, J. E., Lamm, L. O., Haeberli, W.: 
Phys. Rev. C {\bf 50}, 589 (1994)

\bibitem{Ni98}
Nishimori, N., et al.: Nucl. Phys. {A 631}, 697c (1998)

\bibitem{St98}
Stoks, V. G. J.: Phys. Rev. C {\bf 57}, 445 (1998)

\end{thebibliography}
\end{document}